\documentstyle[epsf,prl,twocolumn,aps]{revtex}
\newcommand{\be}{\begin{equation}}
\newcommand{\ee}{\end{equation}}
\newcommand{\bea}{\begin{eqnarray}}
\newcommand{\eea}{\end{eqnarray}}

\newcommand{\la}{\langle}
\newcommand{\ra}{\rangle}

\newcommand{\p}{\partial}
\newcommand{\ri}{{\rm i}}
\newcommand{\re}{{\rm e}}
\newcommand{\rd}{{\rm d}}

\begin {document}
\bibliographystyle {plain}

\title{ Exactly Solvable  Ginzburg-Landau theories  of 
Superconducting Order Parameters  coupled to Elastic Modes
}
\author{Davide Controzzi and  Alexei M. Tsvelik}
\address{Department of Physics, Theoretical Physics, \\
University of Oxford, 1 Keble Road, Oxford, 
OX1 3NP, UK}
\date{\today }
\maketitle

\begin{abstract}
\par
We consider  two families of 
exactly solvable models describing thermal fluctuations in 
two-dimensional superconductors coupled to phonons living in an  
insulating layer,  and
study the stability of the superconducting state with respect to
vortices. The two families are characterized by one or two
superconducting planes. The
results suggest that the 
effective critical temperature increases with  the thickness of the 
insulating layer. Also the presence of the  additional
superconducting layer has the same effect.
\end{abstract}
\sloppy

\par

\section{Introduction}
 
Recently Fateev \cite{fateev1,fateev2} introduced two families 
of integrable models which can
 be interpreted as Ginzburg-Landau free energy functionals 
 describing thermal
 (classical) fluctuations in 
 superconducting films. 
According to  this  interpretation
the models of the first family, which we shall further  call   
type I models, describe a single
superconducting layer deposited on an insulating substrate consisting
of $n$ layers. The
superconducting order parameter interacts with the elastic modes of
 the substrate. The effect of the interaction is to shift 
 the local transition temperature. Type II
 models describe the situation of a double layer where an  insulator is
 sandwiched between two superconducting films. The latter ones may  have
 different critical temperatures. 

These  models provide a rare
opportunity to go beyond weak coupling description  and 
obtain  non-perturbative results 
 for  layered superconductors 
interacting with an  insulating stratum.
 In the case  of   type-I models 
 our principal interest is to find how 
 the Kosterlitz-Thouless transition temperature
 depends on  the thickness of the
 substrate. 
 Type-II models allow us to study 
 the effects of interaction between the 
 superconducting order parameters. 

The paper is organized as follows. In the next Section we describe 
the 
Fateev's models and explain why they can be interpreted    as 
effective Ginzburg-Landau theories  for layered superconductors. In
Section III we study the stability of the type I models with respect to 
the vortices. We do a similar analysis for the type II  models
in Section IV and discuss the results in Section V.

\section{Effective Ginzburg-Landau theory}

In this section we describe  the Fateev's models and construct the
effective Ginzburg-Landau free energy.
The models of the type I
describe a complex bosonic field, 
$\Delta,$ interacting with a $n$-component real scalar
field $\phi=(\phi_1,...,\phi_n)$. There are different models with  slightly
different actions for the scalar fields; 
for our purposes it is sufficient to describe  only 
one of them where the
classical ( Euclidean ) action has the following form:
\bea
\label{model1}
S_{n}^{(I)} &=&\frac{1}{\gamma^2}\int \rd^2 x 
\left \{
2\frac{\partial_{\mu}\Delta \partial _{\mu}\bar\Delta}{1+\Delta\bar\Delta}
+2M_0^2\bar\Delta\Delta \right.
\\ \nonumber 
&&\exp(-\phi_1)+\frac{1}{2}(\partial_{\mu} \phi)^2 
-\frac{M_0^2}{2} \left[ 2\exp(-\phi_1) + \right.
\\ \nonumber
&2&\sum_{i=1}^{n-1} \left.  \left.
\exp(\phi_i-\phi_i+1) + 2 \exp(\phi_n) \right]
\right \}
\eea
This action corresponds to a complex sinh-Gordon model coupled to
affine Toda chains.
The 
complete list of the models of the two families 
is given in \cite{fateev1}.
For small $n$  form (\ref{model1}) 
requires certain  modifications, in particular 
for $n=0$ it becomes\cite{fateev3}:
\be
S_0^{(I)}=\frac{2}{\gamma^2}\int \rd^2 x\left[
\frac{\partial
_{\mu}\Delta \partial _{\mu}\bar\Delta}
{1+\Delta\bar\Delta} +  M_0^2\bar\Delta\Delta\right]
\ee

The type II models describe two complex bosonic fields, $\Delta_{1,2}$,
 interacting
with $n$ elastic modes. Again we present only one specific model where
 the
elastic part coincides with the one described  by Eq.(\ref{model1}). 
The action is the following:
\bea
S_{n}^{(II)} &=&\frac{1}{\gamma^2}\int \rd^2
x\left[ 2
\sum_{s=1,2}
\frac{\partial
_{\mu}\Delta_s \partial _{\mu}\bar\Delta_s}{1+\Delta_s \bar\Delta_s} +
2 M_0^2\bar\Delta_1\Delta_1 \right.
\\ \nonumber
&&\exp(-\phi_1) +2 M_0^2\bar\Delta_2\Delta_2
\exp(-\phi_n)
+\frac{1}{2}(\partial_{\mu} \phi)^2
\\ \nonumber 
-&\frac{M_0^2}{2}&\left. \left \{ 2\exp(-\phi_1)+2\sum_{i=1}^{n-1}
\exp(\phi_i-\phi_{i+1}) + 2 \exp(\phi_n) \right \}
\right]
\eea
For  n=0 the above expression has to be modified as:
\bea
S_{0}^{(II)} &=&\frac{2}{\gamma^2}\int \rd^2
x\left[ \frac{1}{2}
\sum_{s=1,2}
\frac{\partial
_{\mu}\Delta_s \partial _{\mu}\bar\Delta_s}{1+\Delta_s \bar\Delta_s} \right.
 \\ \nonumber
&+& \left. M_0^2 \left\{ \bar\Delta_1\Delta_1+\bar\Delta_2\Delta_2
 +2 (\bar\Delta_1\Delta_1)
(\bar\Delta_2\Delta_2) \right\} \right]
\eea

As a consequence of the $U(1)$ (type I models) or $U(1) \times U(1)$
(type II models)  symmetry, the models have  the
following conserved charges (here we work in Minkovsky space-time):
\be
Q_s=-\frac{2\ri}{\gamma^2}\int \rd
x\frac{(\bar \Delta_s\p_0\Delta_s - \Delta_s\p_0\bar\Delta_s)}
{[1 + 
\Delta_s\bar\Delta_s]}
\ee
where $s=1$ for type I  and $s=1,2$ for type II models. 
In  presence of external chemical potentials $h_s$
the Hamiltonian is modified:  $H = 
H_0 - \sum_s h_s Q_s$. One can introduce  new variables 
\be
\Delta_s  \rightarrow \Delta_s\re^{\ri t h_s}
\ee
to remove the terms with one time derivative. 
Then we obtain the  general form of the Ginzburg-Landau free energy
for our layered superconductors:  
\be
F_n^{(\alpha)}/T = \frac{1}{\gamma^2}\int \rd^2 x {\cal S}_n^{(\alpha)}
\label{freeenergy}
\ee
where:
\bea
\label{f1}
{\cal S}_n^{(I)}&=& 
2 \frac{\partial
_{\mu}\Delta \partial _{\mu}\bar\Delta - h^2\Delta\bar\Delta}{1+ \Delta
\bar\Delta } + 2 M_0^2\bar\Delta\Delta\re^{- 2\phi_1}  \\ \nonumber
&+& \frac{1}{2} (\p_{\mu}\phi)^2 +
2M_0^2\left[\re^{-2\phi_1} + \sum_{i=1}^{n-1}\re^{2(\phi_i - \phi_{i+
1})} + \re^{2\phi_{i}} \right] 
\eea
and
\bea
\label{f2}
{\cal S}^{(II)}_n &=&
\sum_{s=1,2}
2 \frac{\partial
_{\mu}\Delta_s \partial _{\mu}\bar\Delta_s-h_s^2\Delta_s\bar
\Delta_s }{1+\Delta_s
\bar\Delta_s}+\frac{1}{2}(\partial_{\mu} \phi)^2 
\\ \nonumber 
+
2 &M_0^2 &\bar\Delta_1\Delta_1
\exp(-\phi_1) +2 M_0^2\bar\Delta_2\Delta_2
\exp(-\phi_n)   
\\ \nonumber  
-\frac{M_0^2}{2}  && \left [ 2\exp(-\phi_1)+2\sum_{i=1}^{n-1}
\exp(\phi_i-\phi_{i+1}) + 2 \exp(\phi_n) \right ]
\eea
Clearly the fields 
$\Delta_s$ can be interpreted as a superconducting order parameters coupled
to the optical phonon modes $\phi_n$.
We note that the elastic modes are not harmonic,  which is necessary 
to make the model integrable, but in the 
limit $\gamma \rightarrow 0$ we reproduce the conventional
electron-phonon interaction. Indeed,  rescaling the phonon
field $\phi \rightarrow \gamma \phi$,  we  see that the
limit $\gamma \rightarrow 0$ corresponds to harmonic phonons. The fact
that 
 the spectrum of  the Toda chain coincides  with  the spectrum
of harmonic phonons \cite{toda} gives us grounds to  
believe that the unharmonicity present in the model 
does not influence the qualitative validity of our 
results. At the same time it is interesting to have a model with
phononic spectrum that is not restricted to the harmonic one.

In order to get a better understanding of  the nature of the
superconducting state, let us  have a deeper look on  the free
energy for the case $n=0$. 
For the type I models the free energy can  be written as follows:
\be
F_0^{(I)}/T=\frac{2}{\gamma^2}\int \rd^2 x 
\left[
\frac{\partial
_{\mu}\Delta \partial _{\mu}\bar\Delta}
{1+\Delta\bar\Delta} +  V(\bar\Delta,\Delta) \right]
\ee
with
\be
V(\bar\Delta,\Delta)=
M_0^2\bar\Delta\Delta-h^2\frac{\bar\Delta\Delta} 
{1+\Delta\bar\Delta}
\ee

For $M_0 < h$ the effective potential for the order parameter
has a minimum at $|\Delta| \neq 0$ which signal the appearance of
superconductivity. 
Indeed, expanding the action around the minimum we notice
that the quantities $\tau = (M_0^2 - h^2)/\gamma^2$ may  be
interpreted as the distance from the mean field transition 
$(T/T_c - 1)$. Increasing
$h$ one can go from the disordered to the superconducting state. 
One can see it more explicitly using a  semiclassical analysis
  valid for $\gamma \gg 1$. Under the transformation 
\be 
\Delta=\sinh\rho e^{i \varphi}
\label{tr1}
\ee
the free energy becomes:
\be
F_0^{(I)}=\frac{2}{\gamma^2} \int \; d^2x \left \{ 
(\p_\mu \rho)^2 
+\tanh^2\rho(\p_\mu \varphi)^2
+
V^1_{eff}(\rho) \right \}
\label{sc1}
\ee
where:
\be
V_{eff}(\rho,h)=M_0^2 \sinh^2 \rho- h^2\tanh^2\rho
\ee
The measure of integration in the path integral 
changes, but this is  not important for
what we want to discuss here.
For $h/2 > M_0$, $ V_{eff}(\rho,h)$ 
has a minimum at $\rho=\rho_0$, where $\exp{2 \rho_0}=\sqrt{2}(h/M)$. 
Thus when $h$ exceeds the threshold a
gapless Goldstone mode  appears. To see it explicitly we expand the
free  energy around the minimum.  Rewriting $\rho=\rho_0+\xi$ and keeping
only the quadratic terms in $\xi$, we get:
\bea
F_0^{(I)} \sim \frac{2}{\gamma^2} \int \; d^2 x \{(\p_\mu \xi)^2
+(M_0h/2) \xi^2+\tanh^2
\rho_0 (\p_\mu \varphi)^2 \}.
\eea
We can then identify the gapless mode $\varphi$, which  velocity is equal
to the bare one,  and a massive field $\xi$.

A similar analysis can be performed on the type-II models (again we
consider only the simplest case  $n=0$). In this case  the
transformation 
\be
\Delta_s= \sinh\rho_s e^{i\varphi_s}
\label{tr2}
\ee
leads to:
\bea
\label{sc2}
&&F_0^{(II)}=\frac{2}{\gamma^2} \int \; d^2 x \{ 
(\p_\mu \rho_1)^2 +(\p_\mu
\rho_2)^2     \\ \nonumber
&&+\tanh^2\rho_1(\p_\mu \varphi_1)^2 
+\tanh^2 
\rho_2(\p_\mu \varphi_2)^2 
+
V^2_{eff}(\rho_1,\rho_2) \}
\eea
where:
\be
V^2_{eff}(\rho_1,\rho_2)=\sum_s V_{eff}(\rho_s,h_s)+2
M_0^2 
\sinh^2\rho_1
\sinh^2 \rho_2
\ee
This potential can develop one or two minima, depending on the values
of $h_s$, thus  generating either  one or two gapless modes.

From the exact solution one can see that massless modes 
 appear when $h$ exceeds
certain threshold value $M$, where $M$ is a function of the coupling
constant $\gamma$ and parameter $M_0$ ( a similar condition is
obtained for the type II models). In view of the above considerations
these modes can be interpreted
as fluctuations of the superconducting phase and naively one may
identify 
$h = M$ with the onset of superconductivity.

This  interpretation is incorrect however, since it does not take into
account vortices. Massless phases of the Fateev's
models, like any other two-dimensional critical theories with U(1), or
$U(1)\times U(1)$, 
symmetry, may be
unstable with respect to vortices. The reason being that the naive
approach doesn't take into account non-analytic configurations of the
fields, that on a lattice may give a finite contribution to the free energy.
The real transition is of 
the Beresinskii-Kosterlitz-Thouless (BKT) type \cite{bkt} and occurs at
temperature below the mean field transition temperature established by
the condition $h = M$. 
One simple way to see the origin of the vortex configurations of the
order parameter field is the following. Transformations
(\ref{tr1}), (\ref{tr2}) violate an important property of the original
model (\ref{freeenergy}), namely the periodicity of its action. 
The order parameter
fields of the original
models (\ref{f1}), (\ref{f2}) 
 are periodic in $\varphi_s$, while this
periodicity is lost after the transformations (\ref{tr1}), (\ref{tr2}).
To recover
the original periodicity one should add to the forms (\ref{sc1}),(\ref{sc2})
exponents of the dual field (see for instance \cite{gnt}).
These terms
are not  contained in the models we
consider, therefore the latter ones  can provide an adequate
description of the superconducting phase only below the BKT transition
temperature, where vortices are irrelevant. 

In the next sections we
will use the exact solution to study the relevance of vortices in 
our models. To do this we shall need to study more carefully the gapless state
and extract from the Bethe Ansatz equations
the scaling dimensions of the vortex operators. 

\section{The 
Berezinskii-Kosterlitz-Thouless transition in  the monolayer models}

Let us consider the type-I models first. The Bethe ansatz solution deals
with the Minkovsky version of the
models. In our analysis we shall benefit from the fact that the ground
state energy of the (1 + 1)-dimensional field theory with coupling
constant $\gamma$ is equal to the
free energy of the 2-dimensional classical theory with temperature $T
= \gamma^2$: 
\be
F/T={\cal E}
\ee
In the Minkovsky version the appearance of  
the gapless state is related to the creation of    
a condensate of
kinks. Then  using the Bethe ansatz solution one can express 
 the ground state energy
per unit area in terms of  solution of the integral
equation:
\bea
\epsilon(\theta) - \int_{-B}^B \rd\theta'R(\theta -
 \theta')\epsilon(\theta') = M\cosh\theta - h
\label{en}
\eea
\bea
F/T={\cal E} = \frac{M}{2\pi}\int_{-B}^B \rd\theta
 \cosh\theta\epsilon(\theta)
\eea
where $\epsilon(\theta) < 0$ for $|\theta| < B$ and 
the integration limit  $B$ is defined by the condition
\be
\epsilon(\pm B) = 0.
\label{b}
\ee
The kernel $R$ is related to the two-body
 $S$-matrix:
\bea
R(\theta) = \frac{1}{2\pi\ri}\frac{\rd \ln S(\theta)}{\rd \theta}
\eea
According to
\cite{fateev1} its Fourier transform, $R(\omega)$, has the following form:
\bea
1 - R(\omega) = \frac{\sinh[\pi\omega(1 - g)/\alpha]\cosh[\pi\omega(\alpha +
 2g)/2h]}{\cosh(\pi\omega/2)\sinh(\pi\omega/\alpha)} \label{kernel}
\label{kernel1}
\eea
where   
\be
g=\frac{4 \pi}{\gamma^2+4 \pi} \; \; \;
\alpha = \frac{H\gamma^2 + 4\pi
G}{4\pi + \gamma^2}
\ee
with  $G, H$ depending on the model. For model (\ref{model1}) one has
$G = H = 2(n+2)$; $G,H$ always scale with $n$ when $n \rightarrow
\infty$. 
It was shown by Fateev that 
\bea
{\cal E} = - \frac{h^2}{2\pi[1 - R(\omega =0)]}g(h/M)
\eea
where $g(\infty) = 1$. 

Another quantity that will be useful in the following is the dressed
charge $\zeta(\theta)$\cite{bik,book} defined by the following
integral equation:
\be
\zeta(\theta) -  \int_{-B}^B \rd\theta'R(\theta -
 \theta')\zeta(\theta') = 1 \label{zeta}.
\ee
The kernel $R$ is defined in (\ref{kernel1}) and the limit 
$B$ is determined by condition (\ref{b}).

 As we have said, two-dimensional U(1)-symmetric critical points  
can be unstable with respect to the presence of vortex
configurations of the order parameter field. The vortices constitute  
 potentially relevant 
perturbations which appear in the effective action as   exponents
of the dual field (see, \cite{bkt} and, for example ,\cite{gnt}).
This operators become irrelevant if their scaling dimension,
$d_\Theta$, is greater  than two.  
When a  critical theory includes just one U(1) field $\alpha$, the scaling
dimension of the order parameter $\Delta = \exp(i\alpha)$,
$d_{\Delta}$,
is related
to the scaling dimension of the vortex perturbation as follows:
\bea
d_{\Delta}d_{\Theta} = 1/4
\eea
Hence  the
superconducting regime exists   at 
\be
d_\Delta < 1/8.
\label{cond}
\ee

The scaling  dimension for the  primary field $\Phi$ is defined as:
\be
d_\Phi=\Delta^+ +\Delta^-
\ee
where the conformal dimensions $\Delta^\pm$ determine the asimptotics
of the correlation functions of primary fields:
\be
<\Phi_{\Delta^\pm} (x,t) \Phi_{\Delta^\pm} (0,0)>=\frac{\exp(-2 i {\cal D} p_F
x)}
{(x-ivt)^{2 \Delta^+}(x+ivt)^{2\Delta^-}}.
\ee
Here $2 {\cal D}$ 
is the momentum of the state in units of the Fermi momentum $p_F$. 

In (1+1)-dimensional critical theories the scaling dimension of
primary fields  are related to the finite size corrections to
the ground state energy \cite{cardy1,cardy2}. For 
models  which, as
the model in question,  has central charge $c$=1, conformal  dimensions 
are given by:
\be
2\Delta^\pm(\Delta N,D,N^\pm)=2N^\pm+ \left (\frac{\Delta N}{2 {\cal
Z}}
\pm {\cal Z} D \right )^2.
\label{dz}
\ee
The quantum number $\Delta N$ is characteristic of the local field
under consideration; in the context of this model it represent the
number of particles produced by the primary field in
consideration. The quantum numbers $D$ and $N^\pm$, generate the tower
of excited states, and represent respectively the number of particles
that undergo back scattering processes from one Fermi boundary to the
other and the number of particles added at $B (N^+$) or $-B(N^-)$. 
While $\Delta N$ and $D$  are  fixed by
the local field,  $N^\pm$
must be chosen to give the leading asimptotics in the correlation
functions, which is equivalent to minimize $\Delta^\pm$.
In integrable models the quantity ${\cal Z}$ is 
related to the dressed charge introduced above in the
following way:
\be
{\cal Z}=\zeta(B).
\ee

For the order parameter operator we have (see Appendix):
\bea
d_\Delta = \Delta^+(1,0,0)+\Delta^-(1,0,0)=1/[2 {\cal Z}]^2.
\label{sdim}
\eea

In order to study the stability of our models with respect to vortices
we then have to calculate the value of the dressed charge at the Fermi point.
We can make two preliminary observations. 
First, at $B \rightarrow \infty$,
corresponding to $h/M \rightarrow \infty$,  we can approximate
Eq. (\ref{zeta}) by  the Wiener-Hopf (WH) one. Then  we have: 
\bea
d_{\Delta} = \frac{1}{4}[1 - R(\omega = 0)]=
 \frac{1}{4}\frac{\gamma^2}{\gamma^2 + 4\pi}
\eea
This gives us the upper estimate for existence of the
superconductivity:
\be
\gamma^2 < 4\pi, \; \; \; g > 1/2
\ee
This result holds for {\it all} type I models.
Second, since   $R(\theta)$ is a non-singular kernel, 
at small $B$ we have
$\zeta(B) \rightarrow 1$ and the scaling dimension is too large for
the superconductivity to occur. Therefore there is a line  in the
$\gamma -h$-plane separating the superconducting and  the disordered
regions. 

As we noted before for model (\ref{f1}) $G$ and $H$ scale like $n$
for large $n$. Then the kernel (\ref{kernel1}) becomes a delta
function in the 
limit $n \rightarrow
\infty$.
In this case one has to be a bit careful with determining
${\cal Z}$, but the outcome is simple: the scaling dimension is
$h$-independent and is equal to 
\be
d_\Delta = (1 - g)/4 
\ee

We can study  in more 
details the behavior of Eq.(\ref{zeta}) in the regions $B\gg 1$ and
$B\ll 1$ and  obtain the 
asymptotics of the line  separating the superconducting from the
disordered region.
The asymptotic form of the scaling dimension at large $B$ can be found
from the generalized Wiener-Hopf method \cite{foz} and is equal to 
\bea
d_\Delta  &=& \frac{1}{4}(1 - g)[1 + a(M/2h)^{\mu} + ...]\nonumber\\
\mu &=& \frac{2h}{h + 2g}
\eea
At $g - 1/2 << 1$ the condition $d_\Delta < 1/8$ gives us 
\bea
(h/2M) > f(N)(g - 1/2)^{-1/\mu}
\eea
with $\mu$ approaching 2 at $N \rightarrow
\infty$. At $n= 0$ we have $\mu = 2/(1 + g)$.

To calculate ${\cal Z}$ for $B\ll 1$ it is better to use the following
identity:
\be
2 {\cal Z}= \pi v_F \frac{\partial D_\epsilon}{\partial h}
\label{rel}
\ee
where  $D_\epsilon$ is:
\be 
D_\epsilon=\int_{-B}^{B} d\theta \sigma(\theta),
\ee
$v_F$ is the Fermi velocity and $\sigma(\theta)$ 
is the ground state density  for $\epsilon(\theta)$. 
For small $B$ all objects in equation (\ref{rel}) can be calculated
explicitly  and one has 
\be
d_\Delta=\frac{1-g}{8} \left (\frac{M}{h-M} \right )^{\frac{1}{2}}
\ee

We have solved Eq.(\ref{zeta}) numerically for various values of $n$
and constructed the phase diagram of the  model using Eq.(\ref{cond}) to
identify the superconducting region. The results are presented in
Fig.1.
It is clear that the insulating 
substrate   strongly affects the superconducting transition, in
particular with 
increasing $n$ the distance between the BKT transition  temperature and
the mean
field transition temperature  decreases. This means that   
 the effective critical temperature increases with $n$. To our
knowledge this is the first model that displays  such  characteristic
behavior.

\begin{figure}[here]
\epsfxsize=3.5in
\centerline{\epsfbox{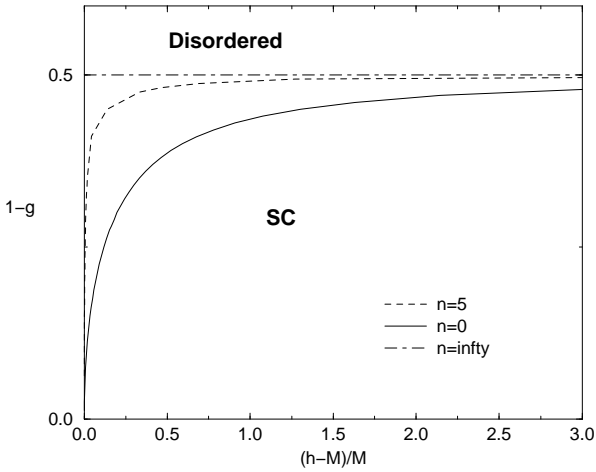}}
\vskip 0.1truein
\protect\caption{Phase diagram of the type I models. The curves
separate the superconducting region from the
disordered one.
}
\label{en1}
\end{figure}

\section{Effect of the interaction between superconducting 
order parameters: the two-layer model}

The situation is considerably more complicated for the case of 
two-layer models. Since we want to focus our  attention on the effects of
the interaction between the order parameters we consider only the $n=0$
version of the model. We have noticed elsewhere 
that in this case the two gapless
modes have the same velocity \cite{us}, and hence  at this point the
model is conformally invariant with the central charge $c$=2.
This result is not general for the Fateev's models but valid only for the
specific case $n=0$. However, the results presented below  can be
generalized for a  case when  the two velocities are
different. In the latter  case the low energy sector is split into two
quasi-independent sub-sectors. The system as a whole is not  
conformally  invariant, 
but each sub-sector is (with its own velocity). Conformal symmetry of
each sub-sector makes it 
 possible to generalize  the basic relations
between the critical exponents and finite size scaling amplitude
\cite{w,ikr} that we will use in the further discussion.

The low-lying excitations for the type-II model with $n=0$ are 
described  by the following  equations:
\bea
\epsilon_1^{(+)}+\left \{ 1-R_1(\omega) \right \}
*\epsilon_1^{(-)}&=&\epsilon_1^0(\theta) 
 \\ \nonumber
&+& \left \{ K_1(\omega)\right \}*[\epsilon_\nu^{(-)}
+\epsilon_{-\nu}^{(-)}] \\
\epsilon_{\pm\nu}^{(+)}+ \left\{
1-R_\nu(\omega) \right\} *\epsilon_{\pm\nu}^{(-)}&=&
\epsilon_{\pm \nu}^0+\left \{ K_1(\omega) \right \}* \epsilon_1^{(-)}
\eea
where we have used the shorthand  notation for convolution:
\be
\left\{ g(\omega) \right\} *f=g*f(\theta)=
\int _{-\infty}^{+\infty} d \theta '
g(\theta-\theta')f(\theta')
\ee
 As usual, 
$\epsilon_{1}^{(+)}(\theta)=\epsilon_{1}(\theta)$ for
$
|\theta |>B$ and zero otherwise,
while $\epsilon_{1}^{(-)}(\theta)=\epsilon_{1}(\theta)$ for
$
|\theta|<B$, and similarly  $\epsilon_{\pm \nu}^{(+)}(\lambda)
=\epsilon_{\pm \nu}(\lambda)$ for $|\lambda|>Q$ and $
\epsilon_{\pm \nu}^{(-)}(\lambda)
=\epsilon_{\pm \nu}(\lambda)$ for $|\lambda|<Q$. This relations
also fix $B$ and $Q$.
The bare energies are:
\be
\epsilon_1^0(\theta)=M \cosh(\theta)-\frac{1}{2}(h_++h_-), \; \;
\epsilon_{\pm\nu}^0=  h_\pm
\ee
and the quantities $h_\pm$ are related to the chemical potentials:
\be
h_\pm=h_1 \pm h_2
\ee
Here
$h_1$ and $h_2$ must be chosen such that $h_\pm$ are always positive.
The Fourier transforms of the kernels have the following form:
\bea
K_1(\Omega)&=&\frac{\sinh \Omega}{\sinh(\nu \Omega)}, \; \; 
R_\nu(\Omega)=\frac{\sinh((\nu-2)\Omega)}{\sinh(\nu \Omega)} \\
\nonumber
R_1(\Omega)&=&1-\frac{\sinh{\Omega} \sinh{\Omega
g}}{\sinh{\frac{(2-g)}{1-g}\Omega} \sinh{\Omega \nu}},  
\; \;\Omega=\frac{\pi \omega}{2 \Lambda}(1-g)
\eea
and the remaining constants are: $\nu^{-1}=1-g, \; \Lambda=2-g$.

For $\frac{1}{2}(h_++h_-) > M$ the mode $\epsilon_1$ becomes gapless
and, depending on the relative values of $h_+$ and $h_-$, it can also
induce a second gapless mode. In the following we will consider 
$h_+ \gg M$. In this
case 
$\epsilon_{\nu}$ is always positive and decouples from the other
equations such that they  become:
\bea
\label{epsilon1}
\epsilon_1(\theta)&=&\epsilon_1^0(\theta)+\int_{-B}^{B} d\theta'
R_1(\theta-\theta') \epsilon_1(\theta ')
\\ \nonumber 
&+& \int_{-Q}^{Q}
d\lambda
K_1(\theta-\lambda)\epsilon_{-\nu}(\lambda)
\\
\label{epsilonnu}
\epsilon_{-\nu}(\lambda)&=&\epsilon_{-\nu}^0 + 
\int_{-Q}^{Q}
d\lambda' R_\nu(\lambda-\lambda')\epsilon_{-\nu}(\lambda')
\\ \nonumber 
&+&
\int_{-B}^{B} d\theta K_1(\lambda-\theta) \epsilon_1(\theta)
\eea
For $h_- \ll h_+$ it is convenient to invert the kernels in
Eq.(\ref{epsilonnu}):
\bea
\label{gs11}
\left \{ \frac{\sinh\nu\Omega}{2\sinh\Omega \cosh[(\nu-1)\Omega]}
\right \}
*\epsilon_{-\nu}^{(+)}+\epsilon_{-\nu}^{(-)}=\nu h_-/2 \\ \nonumber
+\left \{ \frac{1}
{2\cosh[(\nu-1)\Omega]} \right \} *\epsilon_1^{(-)}.
\eea
Using this form in Eq.(\ref{epsilon1}) we get:
\bea
\epsilon_1^{(+)}+K*\epsilon_1^{(-)}=M\cosh\theta-h_-/2-
\\ \nonumber
\left \{
\frac{1}{2\cosh[(\nu-1)\Omega]} \right \}
*\epsilon_{-\nu}^{(+)}
\eea
where
\be
K(\Omega)=\frac{\sinh\Omega \cosh[(3\nu-1)\Omega]}
{2\sinh\nu\Omega 
\cosh[(\nu-1)\Omega]\cosh[(\nu+1)\Omega] }
\ee
Since $\epsilon_{-\nu}^{(+)}$ is very small, to first approximation this
reduces to:
\be
\epsilon_1(\theta)=\epsilon_1^0(\theta)
+\int_{-B}^{B} d\theta'
R(\theta-\theta') \epsilon_1(\theta ')
\ee
where the new kernel $R(\Omega)$ is given by:
\be
R(\Omega)=1-K(\Omega)
\label{rr}
\ee

Clearly 
$\epsilon_{-\nu}$ become gapless for $h_- < h_c$ defined as:
\be
h_c=-\frac{1}{2 \nu^2} \int_{-B}^{B} 
d \theta \frac{\sin{\frac{\pi}{\nu}}}{ \cosh{\frac{\pi
\theta}{\nu}}+\cos{\frac{\pi}
{\nu}}} \epsilon_1(\theta)
\ee

Already at this stage some interesting features emerge. Namely, in
order to get the phase transition on the mean field level, one does
not need both fields $h_1,h_2$ to exceed the critical value for a
single superconductor. Instead of the condition $|h_1| \sim |h_2| \sim
M/2$ we get a weaker condition  $h_1 > M/2$ and $h_2>h_1-h_c$.
 
 Now let us discuss 
how the mean field picture is modified by the  vortices.
For $h_- >  h_c$ when there is only one gapless mode 
we can calculate the scaling dimension of the
order parameter using the procedure of the previous
section. However, for 
$h_- <  h_c$ when there are two gapless modes 
this  procedure should be modified. In this case one needs to introduce
the
dressed charge
matrix \cite{ikr,w,fk},
\be
\label{zgen}
Z=\left( \begin{array}{ll}
{\cal Z}_{11} & {\cal Z}_{1\nu} \\
{\cal Z}_{\nu 1} & {\cal Z}_{\nu \nu}
\end{array} \right )=
\left( \begin{array}{ll}
\zeta_{11}(B) & \zeta_{1\nu}(Q) \\
\zeta_{\nu 1}(B) & \zeta_{\nu \nu}(Q)
\end{array} \right ),
\ee
(we follow the convention introduced in ref.\cite{ikr,fk})
and the function $\zeta(\theta)$  
is determined by the following equations:
\bea
\label{z1}
\zeta_{11}(\theta)&=& 1+\int_{-B}^{B} \rd \theta' R_1(\theta-\theta')
\zeta_{11}(\theta ')
\\ \nonumber
&+&\int_{-Q}^{Q} \rd \lambda K_1(\theta-\lambda)
\zeta_{1 \nu}(\lambda) \\
\zeta_{1 \nu}(\lambda)&=&\int_{-Q}^{Q} \rd \lambda '
R_\nu(\lambda-\lambda ')
\zeta_{1 \nu}(\lambda ') \\ \nonumber 
&+&\int_{-B}^{B} \rd \theta K_1(\lambda-\theta)
\zeta_{11}(\theta) \\
\zeta_{\nu 1}(\theta)&=&\int_{-Q}^{Q} \rd \lambda K_1(\theta-\lambda)
\zeta_{\nu \nu}(\lambda)
\\ \nonumber 
&+&\int_{-B}^{B} \rd \theta R_1(\theta-\theta')
\zeta_{\nu 1}(\theta') \\
\label{z4}
\zeta_{\nu \nu}(\lambda)&=&1+\int_{-B}^{B} \rd \theta K_1(\lambda-\theta)
\zeta_{\nu 1}(\theta )
\\ \nonumber
&+&\int_{-Q}^{Q} \rd \lambda R_\nu(\lambda'-\lambda)
\zeta_{\nu \nu}(\lambda') 
\eea
We note that the structure of this equations is very
similar to the one for the Hubbard model away from half filling, 
where the two gapless modes
correspond to spin and charge excitations \cite{w,fk}.

In this general case the formula for the conformal dimensions are given
by:
\bea
2\Delta_1^\pm({\bf \Delta N,D,N^\pm})=2N_1^\pm+ \left (
{\cal Z}_{11}D_1+{\cal Z}_{\nu 1}D_\nu  \right. \\ \nonumber \pm
 \left. \frac{{\cal Z}_{\nu \nu}\Delta N_1-{\cal Z}_{1 \nu}
\Delta N_\nu}{2 \det Z}
 \right ) ^2 \\
2\Delta_\nu^\pm({\bf \Delta N,D,N^\pm})=2N_\nu^\pm+ \left (
{\cal Z}_{1\nu}D_1+{\cal Z}_{\nu \nu}D_\nu  \right. \\ \nonumber \pm 
\left. \frac{{\cal Z}_{11}\Delta N_\nu-{\cal Z}_{ \nu 1}\Delta N_1}{2 \det Z}
 \right )^2
\eea
The quantum numbers ${\bf \Delta N}=(\Delta N_1,\Delta N_\nu), {\bf
D}=(D_1,D_\nu)$ and ${\bf N^\pm}=(N^\pm_1, N^\pm_\nu)$ are the obvious
generalization of the one defined in the previous section.
For two coupled Gaussian models with a total central charge $c=2$,
the scaling dimensions for primary fields are given by 
\bea
d({\bf \Delta N,D})=
\left (
{\cal Z}_{11}D_1+{\cal Z}_{\nu 1}D_\nu  \right) ^2 +
 \left( \frac{{\cal Z}_{\nu \nu}\Delta N_1-{\cal Z}_{1 \nu}
\Delta N_\nu}{2 \det Z}
 \right )^2 \\ \nonumber
\left (
{\cal Z}_{1\nu}D_1+{\cal Z}_{\nu \nu}D_\nu  \right) ^2 
+ 
\left( \frac{{\cal Z}_{11}\Delta N_\nu-{\cal Z}_{ \nu 1}\Delta N_1}{2 \det Z}
 \right )^2
\eea
When two sectors of the Gaussian model have different velocities 
$v_1$ and $v_2$, 
 the correlation functions of primary fields are
given by:
\bea
&&<\Phi_{\Delta^\pm} (x,t)\Phi_{\Delta^\pm}(0,0)>= \\ \nonumber
&&\frac{ \exp \left[ -2i({\cal D}_1 p_{F1}+{\cal D}_2 p_{F2})x \right] }
{(x-iv_1t)^{2\Delta^+_1}(x+iv_1t)^{2\Delta^-_1}
(x-iv_2t)^{2\Delta^+_2}(x+iv_2t)^{2\Delta^-_2}}
\eea
The vortices generate operators represented by  exponents of the dual
phases of the two gapless fields. We shall  call the scaling dimensions
of this operators $d_{\Theta_1}$ and $d_{\Theta_\nu}$ respectively. 
The conditions
for irrelevance of vortices are :
\be
d_{\Theta_1}>2 \; , \; d_{\Theta_\nu}>2
\label{cond12}
\ee

To see the effect of the interaction between the order
parameters on the stability of the superconducting state with respect
to vortices we shall consider the dependence of the scaling
dimension of the potentially relevant operators  as
a function of $h_-$ for fixed $h_+ \gg M$. 
In particular we will compare the scaling dimensions
in two extreme limits where Eqs (\ref{z1}-\ref{z4}) can be studied
analytically.
The limit of small
$h_-$, which corresponds to the physical situation in which the two
superconductors have the same bare critical temperature, and the limit
$h_- > h_c$.
Having fixed $h_+$ correspond to fix, for example, $h_1$
which is proportional to the mean field critical 
temperature of layer 1. We want
to observe how the presence of the other superconductor affects the
scaling dimension of the order parameter of this superconductor,
i.e. his effective critical temperature.

For $h_- \ll 1$ the matrix problem given by Eqs. (\ref{z1}-\ref{z4}) 
can be reduced by to scalar one by Fourier transforming with respect to
$\lambda$.  
In these  circumstances  the dressed charge matrix has the form:
\be
\label{zmatrix}
Z=
\left( \begin{array}{ll}
\zeta(B) & 0 \\
\zeta(B)/2  & \sqrt{\nu/2}
\end{array} \right )
\ee
where $\zeta(\theta)$ is determined by the following equation:
\be
\zeta(\theta)=1+ \int_{-B} ^{B} \rd \theta' R(\theta-\theta')
\zeta(\theta')
\label{zz}
\ee
and the kernel is defined by Eq.(\ref{rr}).
In this limit the scaling dimension is given by:
\be
d({\bf \Delta N,D})=(\zeta(B) D_1+\frac{\zeta(B)}{2}
D_\nu)^2+\frac{\nu}{2}
D_\nu^2
\ee
and, as shown in Appendix:
\bea
d_{\Theta_1}=d((0,0),(1,0)), \\
d_{\Theta_\nu}=d((0,0),(1,-2)).
\eea
Then in general the model will present two different BKT temperatures.

In the limit  $B>>1$ Eq.(\ref{zz}) can be solved analytically. Using
again the WH method we obtain:
\be
\zeta(B)=\sqrt{2 \nu}
\ee
from which:
\be
d_{\Theta_1}=d_{\Theta_2}=2 \nu=\frac{2}{1-g}
\ee
Then in  this particular limit the conditions
(\ref{cond12}) are always satisfied and the system is
always stable with respect to vortices.
We notice that there are operators characterized by $D_1=1$ and
$D_\nu=-1$ or $D_1=0$  and $D_\nu=1$ with a smaller scaling dimension:
\be
d((0,0),(1,-1))=d((0,0),(0,1))= \nu \equiv d_{12}^\pm
\ee
Again following  the results in the Appendix one can easily see that these
operators are associated with the bosonic  exponents containing  linear combinations of 
 the dual phase   of the
two fields.  These  operators  become
relevant for $g< 1/2$. 

For $h_-> h_c$ one  again has  only one gapless mode and repeating the
procedure of the previous section we get, for $B\gg1 $:
\be
d_{\Theta_1}(h > h_c)=\nu,
\ee
in agreement with the results obtained for the type I models.

Then the effect noticed at the mean field level survives  a deeper
analysis. The effective critical temperature of a superconductor is
enhanced by the presence of another one. The effect is 
present only if the critical temperature of this superconductor is
above some critical value and is maximal  when the two superconductors
have the same critical temperature.

\section{Discussion}

We have studied some properties of two families of integrable 
models describing
two dimensional superconductors interacting with an insulating
substrate. 
The models  show some interesting
features that may be  general properties of layered
superconductors. Due to integrability of the models in question one
can study them in a  strong coupling regime. So  most of the
results presented here are non-perturbative, as suggested by
the previous analysis \cite{us}.

The models presented here can be grouped into  two families.
The members of one family describe a
single superconducting plane interacting with $n$ insulating
layers. This gives a possibility to 
 study the  dependence of the BKT transition temperature on 
$n$. Remarkably the presence of the insulating stratum makes
the system more stable with respect to the BKT transition and
the transition  temperature increases with $n$. 
The models of the second family describe an insulator  sandwiched between
two superconductors, and from this it is possible to extract 
 information about the effect of the
interaction between the superconducting  order parameters.
We have considered only the simplest model of this family, given by
$n=0$. Even in this case  the presence of the other superconductor
stabilizes the system with respect to the BKT transition.
For most of the results presented here it 
is essential  that the systems we consider   are  two
dimensional.  

\section*{Acknowledgements}

We would like to thank H. Frahm for interesting discussions 
and interest to the work.

\section*{Appendix
}
In this appendix we summarize some basic results on the correlation
functions of exponents of bosonic  fields in Gaussian  field theories
and present a simple way to identify the operators
that correspond to  various combinations of 
quantum numbers ${\bf \Delta N} $ and ${\bf D}$.
Let us consider the Gaussian model:
\be
S = \frac{1}{2}\int \rd\tau\rd x[v^{-1}(\p_{\tau}\Phi)^2 + 
v(\p_x\Phi)^2]  
\label{eq:first}
\ee
It is a well known  that the correlation function of exponents of bosonic 
 fields is given by:
\bea
\label{eq:exponents}
\la\exp[\ri\beta_1\Phi(\xi_1)]...\exp[\ri\beta_N\Phi(\xi_N)]\ra = 
\\ \nonumber
\prod_{i > j}\left(\frac{z_{ij}\bar
z_{ij}}{a^2}\right)^{(\beta_i\beta_j/4\pi)}
\left(\frac{R}{a}\right)^{ - \left(\sum_n\beta_n\right)^2/4\pi} 
\eea
where $z = \tau + \ri x/v, \bar z = \tau - \ri x/v$ and in an infinite
system  this is different from zero only if:
\be
\sum_n\beta_n = 0
\ee

The expression (\ref{eq:exponents})
is factorized into analytic and anti-analytic parts
and then it is useful to rewrite it as:
\bea
\la\exp[\ri\beta_1\Phi(\xi_1)]...\exp[\ri\beta_N\Phi(\xi_N)]\ra\nonumber\\
= G(z_1,...,z_N)G(\bar z_1,...,\bar z_N)\delta_{\sum\beta_n,0} 
\label{eq:corrfunc}
\eea
where
\[
G(\{z\}) = \prod_{i >
j}\left(\frac{z_{ij}}{a}\right)^{(\beta_i\beta_j/4\pi)}
\]
This factorization guarantees that  analytic and anti-analytic parts of
the correlation functions can be  studied independently. 
Since factorization of the 
correlation functions is a general fact, it  can be 
formally written as factorization of the corresponding fields. 
Then inside  the $\la...\ra$-sign   one 
can rewrite $\Phi(z,\bar z)$ as a sum of  independent analytic and 
anti-analytic fields:
\bea
\Phi(z,\bar z) = \phi(z) + \bar\phi(\bar z),\\
\exp[\ri\beta\Phi(z,\bar z)] = \exp[\ri\beta\phi(z)]\exp[\ri\beta\bar\phi(\bar 
z)]
\eea

For many purposes it is convenient to introduce 
 the `dual' field $\Theta(z,\bar z)$
defined  as 
\be
\Theta(z,\bar z) = \phi(z) - \bar\phi(\bar z)
\ee
which   satisfies the 
following equations:
\be
\p_{\mu}\Phi = - \ri\epsilon_{\mu\nu}\p_{\nu}\Theta \label{eq:Couchy}
\ee

In order to  study correlation functions of the analytic and the 
anti-analytic fields, we define  the fields 
\bea
A(\beta,z) \equiv  
\exp\left\{\frac{\ri}{2}\beta[\Phi(z,{\bar z}) + \Theta(z,{\bar 
z})]\right\}\nonumber\\
\bar{A}(\bar\beta,\bar{z}) \equiv  
\exp\left\{\frac{\ri}{2}\bar{\beta}[\Phi(z,{\bar z}) - \Theta(z,{\bar 
z})]\right\}
\eea
with, generally speaking,   different $\beta,\bar\beta$.  With the 
operators $A(\beta,z),\bar{A}(\bar\beta,\bar{z})$  
one  can expand local functionals of mutually nonlocal fields 
$\Phi$ and $\Theta$. 
Suppose that 
$F(\Phi,\Theta)$
is a local functional periodic both in  $\Phi$ and $\Theta$ 
with the periods 
$T_1$ and $T_2$, respectively. This functional can be expanded in terms of the 
bosonic exponents:
\bea
F(\Phi,\Theta)& =& \sum_{n,m}\tilde{F}_{n,m}\exp[(2\ri\pi n/T_1)\Phi + 
(2\ri\pi m/T_2)\Theta] \nonumber\\
&=&\sum_{n,m}\tilde{F}_{n,m}A(\beta_{nm},z)\bar{A}(\bar\beta_{nm},\bar{z})
\label{expansion}
\eea
where
\bea 
\beta_{nm} = 2\pi\left( \frac{n}{T_1} + \frac{m}{T_2}\right) \\
\bar\beta_{nm} = 2\pi\left( \frac{n}{T_1} - \frac{m}{T_2}\right) 
\eea
 It turns out that the periods 
$T_1,T_2$ are not arbitrary, but  related to each
other. The
reason for this lies in the fact that the correlation 
functions must be uniquely defined  on the complex plane. We can see 
how this argument works using the pair correlation function as an
example:
\bea
\la A(\beta_{nm},z_1)\bar{A}(\bar\beta_{nm},\bar{z}_1)A(-\beta_{nm},z_2)
\bar{A}(-\bar\beta_{nm},\bar{z}_2)\ra \nonumber\\
= (z_{12})^{- \beta^2_{nm}/4\pi}(\bar{z}_{12})^{-\bar\beta^2_{nm}/4\pi}
= \frac{1}{|z_{12}|^{2d}}\left(\frac{z_{12}}{\bar{z}_{12}}\right)^{S} 
\label{eq:branch}
\eea
where we introduce the quantities 
\[
d = \Delta^+ + \Delta^- = \frac{1}{8\pi}(\beta^2 + \bar\beta^2)
\]
and 
\[
S = \Delta^+ - \Delta^- = \frac{1}{8\pi}(\beta^2 - \bar\beta^2)
\]
which are called the `scaling dimension' and the 
`conformal spin', respectively.

 The two branch cut singularities in Eq. (\ref{eq:branch}) 
cancel each other and give a
uniquely 
defined  function only if
\be
2S = \mbox{(integer)}
\ee
i.e., physical fields with uniquely defined 
correlation functions must have integer or half-integer conformal
spins. This equation  suggests the relation  
\be
T_1 = \frac{4\pi}{T_2} \equiv \sqrt{4\pi K}\label{eq:periods}
\ee
as the minimal solution.
Here we introduce the Luttinger liquid parameter  $K$ for 
future convenience. The normalization is such that at $K = 1$ the
periods for the field $\Phi$ and its dual  are equal. 
The quantities
$\Delta^+, \Delta^-$  are called 
`conformal dimensions' or `conformal weights'. 
In the  case of the Gaussian model
(\ref{eq:first}) the  
conformal dimensions of 
the basic operators are given  by:
\bea
\Delta_{nm}^+ \equiv  \beta^2_{nm}/8\pi = 
\frac{1}{8}\left(m \sqrt K + 
\frac{n}{\sqrt K}\right)^2\nonumber\\
\Delta_{nm}^- \equiv  \bar\beta^2_{nm}/8\pi = 
\frac{1}{8}\left(m \sqrt K - 
\frac{n}{\sqrt K}\right)^2\nonumber\\
\label{eq:dimensions}
\eea

This equations can be rewritten in terms of the quantum numbers
$\Delta N$ and $D$ previously introduced as:
\be
2\Delta_{nm}^\pm=\left(\frac{D \sqrt K}{2} \pm 
\frac{\Delta N}{ \sqrt K}\right)^2
\label{delta2}
\ee

In integrable models the parameter $K$ is related to the dressed
charge introduced in Sec.III:
\be
K=4 \zeta(B)^2
\ee
from which we obtain the form (\ref{delta2}):
\be
\label{dz}
2\Delta^\pm =\left (\zeta(B) D  \pm\frac{\Delta N}{2 \zeta(B)}
\right )^2.
\ee
Comparing it with equation (\ref{expansion}) we can easily see that the
scaling dimension $d_\Phi$ 
of the field $\Phi$ is given by $\Delta N=1$ and $D=0$, while the one
of the dual field, $d_\Theta$, by $\Delta N=0$ and $D=1$:
\bea
d_\Phi=1/4 \zeta(B)^2 \\
d_\Theta=\zeta(B)^2.
\eea

All the considerations above can be generalized to the case of $n$
gaussian fields with the same velocity:
\be
S = \frac{1}{2}\int \rd\tau\rd x \sum_{i=1}^n
[v^{-1}(\p_{\tau}\Phi_n)^2 + 
v(\p_x\Phi_n)^2]  \label{eqn:first}
\ee
In a similar fashion as before we can introduce the fields:
\bea
\label{an}
A({\bf B},z) \equiv  
\exp\left\{\frac{\ri}{2}{\bf B} \cdot [{\bf \Phi}(z,{\bar z}) + {\bf 
\Theta} (z,{\bar z})]\right\} \\ \nonumber
\bar{A}(\bar {\bf B},\bar{z}) \equiv  
\exp\left\{\frac{\ri}{2} \bar {\bf B} \cdot [{\bf \Phi}(z,{\bar z}) 
- {\bf \Theta}(z,{\bar z})]\right\}
\eea
where ${\bf \Phi}=(\Phi_1,...,\Phi_n)$, ${\bf \Theta}={\Theta_1,...,\Theta_n}$,
${\bf B}=(\beta_1,...,\beta_n)$ and
$\bar{\bf B} =(\bar\beta_1,...,\bar\beta_n)$. The generalization of the
conformal dimension and spin are:
\bea
\nonumber
d({\bf \Delta N,D})&=&1/4{\bf \Delta N}^T{\bf X}^{-1} {\bf \Delta N}+
{\bf D}^T{\bf X} {\bf D} \equiv {\bf \Delta^+ +\Delta^-} \\ \nonumber
S({\bf \Delta N,D})&=&{\bf \Delta N}^T{\bf D} 
\eea 
where ${\bf \Delta N}=(\Delta N_1,...,\Delta N_n)^T$, ${\bf
D}=(D_1,...,D_n)^T$ 
and ${\bf X}=Z^T Z$ with $Z$ is the dressed charge matrix 
defined by (\ref{zgen}). 
Then the  conformal dimension of each field is given by:
\be
\Delta_i^+=\frac{1}{8 \pi} \beta_{i}^2 \; ; \; \Delta_i^-
=\frac{1}{8 \pi} \bar\beta_{i}^2 
\label{deltan}
\ee
For the case of two gapless modes of interest in the paper the forms
(\ref{an}) and (\ref{deltan}) reduces to 
\bea
\label{two}
A({\bf B},z)\bar A(\bar {\bf B},\bar z)
=  \\ \nonumber
\exp \left [ i(Z_{11}D_1 + Z_{21} D_2) \Theta_1+
i(Z_{22} \Delta N_1-Z_{12} \Delta N_2)  \Phi_1 \right. \\ \nonumber 
+ \left. i(Z_{12} D_1+Z_{21} d_2) \Theta_2 +
i (Z_{11} \Delta N_2-Z_{21} \Delta N_1) \Phi_2 \right ]
\eea
It is easy to read from 
(\ref{two}) that the exponent of the dual field $\Theta_1$ is
characterized by $Z_{12} D_1+Z_{21}
D_2=0$
and $\Delta N_{1,2}=0$, while for $\Theta_2$ you need  
$Z_{11}D_1 + Z_{21} D_2=0$
and $\Delta N_{1,2}=0$.
Then for a dressed charge matrix of 
the form (\ref{zmatrix})  the conformal  dimensions  for exponents of 
the two dual fields are
\bea
d_{\Theta_1}&=&d((0,0),(1,0)) \\
d_{\Theta_2}&=&d((0,0),(-1,2)).
\eea


\begin{thebibliography}{99}
\bibitem{fateev1} V. A. Fateev, Nucl. Phys. B{\bf 479}, 594 (1996).
\bibitem{fateev2} V. A. Fateev, Nucl. Phys. B{\bf 473}, 509 (1996).
\bibitem{fateev3} V.A.Fateev, Phys. Lett B {\bf 357},397 (1995).
\bibitem{toda}A.E.Arishtein, V.A.Fateev, and A.B Zamolodchikov,
Phys.Lett. {\bf B 87}, 389 (1979).
\bibitem{bkt}V.L.Beresinskii, Sov.Phys. JEPT {\bf 32} 493 (1971);
J.M.Kosterlitz and D.J.Thouless, J. Phys. C{\bf 6}, 1181 (1973);
J.M.Kosterlitz, J. Phys. C {\bf 7}, 1046 (1974).
\bibitem{gnt}A.O.Gogolin, A.A.Nersesyan and
A.M.Tsvelik, \\ {\em Bosonization and Strongly Correlated Systems},
Cambridge University Press (1999)
\bibitem{belavinetalt}A.A.Belavin, A.M.Polyakov and A.B.Zamolodchikov,
Nucl.Phys. {\bf B270}, 333 (1984).
\bibitem{cardy1}J.L.Cardy, Nucl.Phys. B {\bf 270},186 (1986).
\bibitem{cardy2}H.W.Blote,
J.L.Cardy and M.P.Nightingale \prl {\bf 56},742 (1986); I.Affleck,
{\em ibidem} {\bf 56}, 746(1986).
\bibitem{foz}V. A. Fateev, E.Onofri, Al. A. Zamolodchikov,
Nucl.Phys. B{\bf 406},521 (1993).
\bibitem{us}D.Controzzi and A.M.Tsvelik, Nucl.Phys.{\bf B} 572, 609.
\bibitem{bik}N.M.Bogoliubov, A.G.Izergin and V.E.Korepin, Nucl.Phys. B
{\bf 275}, 687 (1986).
\bibitem{book}V.E.Korepin, N.M.Bogoliubov, A.G.Izergin, {\em  Quantum
Inverse Scattering Method and Correlation Functions}, Cambridge University
Press (1993).
\bibitem{ikr}A.G.Izergin, V.E.Korepin and N.Yu Reshetikhin, J.Phys A
{\bf 22},2616 (1989). 
\bibitem{w}F.Woyanarivich, J.Phys. A {\bf 22}, 4243 (1989).
\bibitem{fk}H.Frahm and V.Korepin, \prb {\bf 42},10553 (1990).
\end{thebibliography}
\end{document}